\documentclass[aps,prb,twocolumn,superscriptaddress,amsmath,amssymb,floatfix]{revtex4}
\usepackage{graphicx}% Include figure files
\usepackage{dcolumn}% Align table columns on decimal point
\usepackage{bm}% bold math

\newcommand{\pdag}{{\phantom{\dagger}}}
\newcommand{\ybco}{YBa$_2$Cu$_3$O$_{y}$}
\newcommand{\ndlsco}{La$_{1.6-x}$Nd$_{0.4}$Sr$_x$CuO$_{4}$}
\newcommand{\eulsco}{La$_{1.8-x}$Eu$_{0.2}$Sr$_x$CuO$_{4}$}
\newcommand{\lsco}{La$_{2-x}$Sr$_x$CuO$_{4}$}

\newcommand{\Tc}{T_{\rm c}}

\newcommand{\bk}{{\vec k}}

\newcommand{\Qch}{{\vec Q}_{\rm c}}
\newcommand{\Qchx}{{\vec Q}_{{\rm c}x}}
\newcommand{\Qchy}{{\vec Q}_{{\rm c}y}}

\newcommand{\Qspx}{{\vec Q}_{{\rm s}x}}
\newcommand{\Qspy}{{\vec Q}_{{\rm s}y}}

\newcommand{\e}{\epsilon}

\begin{document}

\title{Stripe order and quasiparticle Nernst effect in cuprate superconductors}

\author{Andreas Hackl}
\affiliation{Department of Physics, California Institute of Technology, Pasadena, CA 91125, USA}
\author{Matthias Vojta}
\affiliation{Institut f\"ur Theoretische Physik, Universit\"at zu K\"oln,
Z\"ulpicher Stra{\ss}e 77, 50937 K\"oln, Germany}

\date{August 20, 2010}

%%%%%%%%%%%%%%%%%%%%%%%%%%%%%%%%%%%%%%%%%%%%%%%%%%%%%%%%%%%%%%%%%%%%%%%%%%%%%%%

\begin{abstract}
After a brief review of current ideas on stripe order in cuprate high-temperature
superconductors, we discuss the quasiparticle Nernst effect in the cuprates, with focus
on its evolution in non-superconducting stripe and related nematic states. In general, we
find the Nernst signal to be strongly enhanced by nearby van-Hove singularities and
Lifshitz transitions in the band structure, implying that phases with translation
symmetry breaking often lead to a large quasiparticle Nernst effect due to the presence
of multiple small Fermi pockets. Open orbits may contribute to the Nernst signal as well,
but in a strongly anisotropic fashion. We discuss our results in the light of recent
proposals for a specific Lifshitz transition in underdoped \ybco\ and make predictions
for the doping dependence of the Nernst signal.
\end{abstract}

\maketitle

%%%%%%%%%%%%%%%%%%%%%%%%%%%%%%%%%%%%%%%%%%%%%%%%%%%%%%%%%%%%%%%%%%%%%%%%%%%%%%%

\section{\label{intro} Introduction}

Cuprate high-temperature superconductors continue to hold unsolved puzzles, one of them
being the origin of the so-called pseudogap regime.\cite{timusk} Various explanations
have been proposed for this apparent suppression of spectral weight, which occurs below
the doping-dependent pseudogap temperature $T^\ast$. These proposals include
phase-incoherent Cooper pairing, symmetry-breaking orders competing with
superconductivity, exotic fractionalized states, and short-range singlet correlations as
precursor to the half-filled Mott insulator.\cite{norman_rev,leermp}

Stripe order\cite{pnas,stripe_rev1,antonio_rev,stripe_rev2} takes a prominent role in the cuprate
phenomenology: Such periodic modulations of spin and charge densities have been
established to exist in the \lsco\ (or 214) family of cuprates, using a variety of
experimental techniques, most importantly neutron and x-ray scattering.\cite{jt95}
Signatures of stripes also appear in experiments on other cuprate families, albeit
providing less conclusive evidence, either because the experimental probes are restricted
to the sample surface (e.g. scanning tunneling microscopy
\cite{kapi03a,yazdani04,kohsaka07}) or stripe physics is only probed indirectly (e.g. via
finite-energy spin fluctuations, phonon anomalies, or quantum
oscillations).\cite{stripe_rev2} Collectively, the observations suggest that the tendency
toward stripe order is a common phenomenon in underdoped cuprates. Then, while in some
cuprates the stripe order is strong and static, in others it is presumably fluctuating in
space and time, with the possibility of being pinned by impurities.\cite{stripe_rev1,antonio_rev,stripe_rev2}
%Therefore, fluctuating stripes provide an appealing (but at the same time
% somewhat elusive concept,

An interesting probe of pseudogap physics is provided by the transverse thermoelectric
response, also known as Nernst effect. While typically small in conventional metals,
large positive Nernst signals are known to arise from the motion of vortices in type-II
superconductors.\cite{maki,vortex} In addition, it has been pointed out that a large
Nernst signal (of either sign) can occur in metals with a small (effective) Fermi
energy.\cite{nernst_fl} Nernst measurements in cuprates, showing a rise of the Nernst
signal at temperatures above the superconducting $\Tc$, have been commonly interpreted as
evidence for fluctuating Cooper pairs above $\Tc$.\cite{xu00,wang06} However, this Nernst
onset temperature appears to lie below the pseudogap temperature $T^\ast$ identified by
other probes, indicating that fluctuating Cooper pairs do {\em not} account for all of
the cuprate pseudogap.

Recent Nernst measurements on underdoped cuprates have revealed additional information:
In stripe-ordered \ndlsco\ the temperature dependence of the Nernst signal shows an
additional (positive) peak or shoulder at intermediate temperatures which was tentatively
attributed to a Fermi-surface reconstruction due to density-wave order.\cite{taill09}
(Note that this interpretation has been questioned by others.\cite{li10}) In de-twinned
crystals of \ybco\ (YBCO), the normal-state Nernst signal was found to be
negative,\cite{daou09} and to display a huge temperature-dependent in-plane
anisotropy.\cite{taill10b} The latter fact appears to tie in with the tendency to
electron-nematic order, previously identified in neutron-scattering measurements on
YBCO-6.45.\cite{hinkov08}

These measurements underline the importance of understanding the quasiparticle Nernst effect in
cuprates. Theoretically, the Nernst response has been calculated in simple quasiparticle
models, and it has been shown that density-wave order can indeed lead to an enhanced
Nernst signal,\cite{oga04,tewari09,hackl09a,hackl10,zhang09} with the sign depending on the
spatial periodicity and other details of the ordering pattern.\cite{hackl10} Further,
Fermi-surface distortions as expected for electron-nematic order in the $d$-wave channel
have been shown to induce huge Nernst anisotropies,\cite{hackl09b} in semi-quantitative
agreement with the experimental data.\cite{taill10b}
These results suggest (together with successful descriptions of transport and quantum
oscillation data using related models\cite{millis07,sudip,linmillis,harrison09,millis10}) that scenarios
of Fermi-liquid-like quasiparticles, moving in states with lattice symmetry breaking,
capture
some
important properties of underdoped cuprates.

The purpose of the present paper is twofold. First, we briefly review current ideas on
the origin of stripe order in cuprates, together with its effect on the fermionic
quasiparticle spectrum. Nematic order as a precursor of uni-directional stripes will
naturally appear in the discussion. Second, we summarize and extend theoretical results
for the quasiparticle Nernst effect in the normal state of stripe-ordered
cuprates,\cite{hackl10} also discussing the anisotropy of the Nernst signal. In
particular, we focus on the Lifshitz transition where electron pockets near $(\pi,0)$ disappear as
doping is reduced, as recently proposed\cite{millis10} for underdoped \ybco\ on the basis
of quantum oscillation experiments.\cite{sebastian10b}
While small Fermi pockets induce a strongly enhanced Nernst
signal in both directions, open orbits appreciably contribute to the Nernst response only
for a temperature gradient applied parallel to the dominant hopping direction.
This allows specific predictions for the normal-state Nernst signal across the Lifshitz
transition.

\subsection{Outline}

The body of this paper is organized as follows:
In Sec.~\ref{sec:stripe} we start with an overview on selected aspects of stripe physics
which we consider relevant for the Nernst effect and its evolution within the cuprate
phase diagram.
Sec.~\ref{sec:nernst} contains a brief general discussion of thermoelectric effects with
focus on the quasiparticle Nernst signal.
Sec.~\ref{sec:model} then describes the concrete model and calculational scheme which we
employ to investigate the quasiparticle Nernst effect in the normal state of cuprates.
The results from this approach are discussed in Sec.~\ref{sec:results}, where we both
review selected earlier results of Refs.~\onlinecite{hackl10,hackl09b} and present new
ones which are relevant for the strongly underdoped regime.
A discussion of experimental implications and an outlook close the paper.

%%%%%%%%%%%%%%%%%%%%%%%%%%%%%%%%%%%%%%%%%%%%%%%%%%%%%%%%%%%%%%%%%%%%%%%%%%%%%%%

\section{Stripe order in cuprates}
\label{sec:stripe}

Density-wave order in cuprates has been the subject of numerous review articles in the
past,\cite{pnas,stripe_rev1,antonio_rev,stripe_rev2} but new experiments continue to shape (and sometimes
change) our view on this fascinating set of phenomena. We shall use this section to give
a brief overview, combining experimental and theoretical aspects, regarding two central
aspects, namely the cause of stripe formation and the nature of the fermionic
single-particle excitations in stripe phases.

\subsection{Order parameters and symmetry breaking}

To set the stage, we repeat the standard definitions of order parameters which are
relevant for uni-directional stripe order in quasi-2d systems such as cuprates. In all
cases, we employ a language appropriate for slowly varying order parameter fields.

First, a charge density wave (CDW) is described by a pair of complex scalar fields $\phi_{cx}$,
$\phi_{cy}$ for the two CDW directions with wavevector $\Qchx$ and $\Qchy$.
The charge density is assumed to obey
\begin{equation}
\langle \rho ({\vec R}, \tau) \rangle = \rho_{\rm avg} + \mbox{Re} \left[e^{i \Qch
\cdot {\vec R}} \phi_c (\vec R, \tau)  \right] \,.
\label{chargemod}
\end{equation}

The spin density wave (SDW) is assumed to be collinear.\cite{colli_foot}
Then, a pair of complex vector fields $\phi_{s\alpha x}$, $\phi_{s\alpha y}$,
$\alpha=x,y,z$, captures SDW correlations with wavevectors $\Qspx$ and $\Qspy$, with the
spin density following
\begin{equation}
\langle S_\alpha ( {\vec R},\tau ) \rangle = Re \bigl[ e^{i{\vec Q}_s\cdot {\vec R}} \phi_{s\alpha} ({\vec R},\tau )
\bigr].
\label{sdw}
\end{equation}
Experimentally, order has been found at $\Qspx = 2\pi(0.5\pm 1/M,0.5)$,
$\Qspy = 2\pi(0.5,0.5\pm 1/M)$ and $\Qchx = (2\pi/N,0)$, $\Qchy = (0,2\pi/N)$, where $M$
and $N$ are the real-space periodicities which follow $M=2N$ to a good
accuracy.\cite{stripe_rev1,antonio_rev,stripe_rev2,jt95}

As argued in Ref.~\onlinecite{KFE98}, introducing a separate order parameter for
rotational symmetry breaking in a tetragonal environment is useful: this is an Ising
scalar $\phi_n$ for $l=2$ spin-symmetric electron-nematic order which carries wavevector
$\vec Q=0$. It is common practice to refer to $\phi_n$ as nematic order parameter. Care
is required in the terminology: strictly speaking, a nematic phase is one with $\phi_n\neq 0$ and
$\phi_c=\phi_s=0$; phases with e.g. $\phi_c\neq 0$ have $\phi_n\neq 0$ as well, but are
not to be called nematic (but smectic).

In a number of cuprates, the tetragonal in-plane symmetry is broken down to orthorhombic,
and spontaneous nematic order cannot exist. It may still make sense to discuss the
tendency toward electron-nematic order if the microscopic anisotropy is strongly enhanced
by electronic correlation effects.

\subsection{Origin of stripes}

The driving force for stripe formation has been intensively discussed since stripes have
been discovered. Conceptually, weak-coupling and strong-coupling approaches need to be
distinguished.

On the one hand, weak coupling refers to density-wave instabilities which can be
obtained in perturbation theory with a Fermi-liquid picture. In the simplest case, RPA is
used for a given dispersion of single-particle excitations to obtain the ordering
wavevector from the pole in the static susceptibility. This concept relies on the
existence of well-defined quasiparticles with significant weight and ties the ordering
to properties of the Fermi surface.

On the other hand, strong-coupling ideas come in various flavors, most often they do not
relate to properties of the single-particle sector. Popular lines of thought are (A)
frustrated phase separation, (B) spin-charge ``topological'' properties, and (C)
valence-bond solid formation, all of which have been invoked to rationalize the formation
of conducting stripes in cuprates. While (A) is general, (B) and (C) refer to more
microscopic aspects.

The idea of frustrated phase separation\cite{ek93,sk1} builds on the assumption that, without long-range
Coulomb repulsion, a doped Mott insulator is unstable toward phase separation, i.e., the
Mott insulator tends to expel holes. Then, including Coulomb repulsion leads to domain
formation, in the simplest case these domains are linear. (More generally, particles on a
lattice, moving under the influence of short-range attractions and long-range repulsions,
minimize their energy by forming domains with modulated density.) It must be noted,
however, that the existence of phase separation in the Hubbard or $t$--$J$ models in the
relevant regime of parameters is questionable, considering a body of numerical results
where phase separation is only seen in the $t$--$J$ model at large $J/t$ and small doping.

The co-existence of spin and charge order, with hole-rich stripes being anti-phase domain
walls of the underlying antiferromagnetism, has led to interpretations of stripes as a
two-dimensional generalization of holons.\cite{zaanen96} This concept suggests that
magnetic order is a prerequisite for stripe formation; however, at least some
compounds display a temperature range with long-range charge order but without long-range
magnetic order.

Finally, the valence-bond solid idea is based on the tendency of paramagnetic Mott
insulators on the square lattice to break translation and rotation symmetry in the form
of a columnar valence-bond solid.\cite{ssrmp} Upon doping, stripes are assumed to
inherit this type of symmetry breaking, with an additional modulation in the hole
density.\cite{vs99}

To date, the question about the origin of stripes is not settled. Many cuprate
experiments point toward a strong-coupling picture as being more appropriate than a
weak-coupling one.\cite{stripe_rev1,antonio_rev,stripe_rev2} One central argument here is about the
evolution of the charge-ordering wavevector with doping $x$ in 214 compounds, $\Qch
\approx (4\pi x,0)$ for $x<1/8$, which is opposite to the evolution of possible
Fermi-surface nesting wavevectors; others are about the intensity of the magnetic
response (too large to be compatible with RPA) and the pinning of stripes by impurity
doping (which is not expected in weak coupling). Which of the strong-coupling ideas is
closest to reality is not obvious. Recent STM experiments,\cite{kohsaka07} showing
bond-centered stripes with strong bond modulations on the surface of two different
cuprates, lend some support to the valence-bond solid idea, although a detailed
understanding of the role of oxygen orbitals is lacking.

\subsection{Fermionic quasiparticles in stripe phases}

Low-temperature dc transport is generically determined by the properties of the
low-energy charge carriers. Various low-temperature experiments, most notably those
involving quantum oscillations, indicate that charge carriers in underdoped cuprates are
Fermi-liquid-like quasiparticles.\cite{sebastian10,ramshaw10} However, the temperature evolution of
both thermodynamic and spectral properties does not easily fit into a conventional
Fermi-liquid picture.\cite{timusk}
A generally accepted solution of this puzzle is not known to date.

The available proposals can be grouped in two classes:
(i) The low-doping state is asymptotically a conventional Fermi
liquid (in the absence of superconductivity), but with both coherence temperature and
quasiparticle weight being small. This implies, e.g., that the Fermi surface of this state
is ``large'' (in the absence of translational symmetry breaking), i.e., it fulfills
Luttinger's theorem. This large Fermi surface may be unobservable if translational
symmetry breaking sets in above the coherence temperature.
(ii) The low-doping state is a metallic non-Fermi liquid. It may still feature
Fermi-liquid-like quasiparticles, but Luttinger's theorem is violated. (A state of this
type is the so-called fractionalized Fermi liquid, introduced for two-band models of
heavy-fermion metals.\cite{flst})
Theories of type (ii) find support in computational studies of the one-band Hubbard model
using cluster extensions of dynamical mean-field theory.\cite{haule07,jarrell08,civelli08,imada08}
These studies indicate that the
low-doping metallic state is characterized by small Fermi pockets (in the absence of symmetry
breaking), and a transition to a conventional metal with a large Fermi surface occurs
around optimal doping.
A phenomenological ansatz for the self-energy describing a pseudogap state with small
pockets has been put forward in Ref.~\onlinecite{YRZ}.

In both scenarios, density-wave order will lead to a backfolding of the quasiparticle
bands due to translational symmetry breaking. This invariably causes the existence of
multiple Fermi pockets and/or open orbits, depending on the strength of the density-wave
order and other microscopic details.

To date, photoemission experiments have not been able to distinguish the available
scenarios, mainly because of insufficient energy and momentum resolution
(although progress has been made recently\cite{fournier10}).
On the theory
side, many papers start from the more conventional scenario (i) (often even approximating
the quasiparticle weight by unity) which then allows to perform concrete calculations of
more complicated observables with moderate effort. Our calculations below will follow
this route as well.

%%%%%%%%%%%%%%%%%%%%%%%%%%%%%%%%%%%%%%%%%%%%%%%%%%%%%%%%%%%%%%%%%%%%%%%%%%%%%%%

\section{Nernst effect}
\label{sec:nernst}

The Nernst effect is the generation of a transverse electric field by a longitudinal
thermal gradient in the presence of a finite magnetic field. Although the Nernst effect
in cuprate superconductors has attracted some attention over the past years, the Nernst
effect in correlated-electron metals in general is largely unexplored.\cite{nernst_fl}

In linear-response theory, the thermoelectric response is captured by three conductivity tensors
$\hat{\sigma}$, $\hat{\alpha}$, and $\hat{\kappa}$, which relate charge current $\vec{J}$
and heat current $\vec{Q}$ to electric field $\vec{E}$ and thermal gradient
$\vec{\nabla }T$:
\begin{equation}
\left(
\begin{array}{c}
\vec{J} \\
\vec{Q} \\
\end{array}
\right)
=
\left(
\begin{array}{cc}
\hat{\sigma}  & \hat{\alpha} \\
T\hat{\alpha} &  \hat{\kappa}\\
\end{array}
\right)
\left(
\begin{array}{c}
\vec{E} \\
-\vec{\nabla} T\\
\end{array}
\right).
\label{thermoelectrics}
\end{equation}
The electrical field induced by a thermal gradient in the absence of an
electrical current can be expressed by the linear-response relation
$\vec{E}= -\hat{\vartheta } \vec\nabla T$, and
Eq. \eqref{thermoelectrics} together with $\vec{J}=0$
yields $\vec{E}=\hat{\sigma}^{-1} \hat{\alpha} \vec\nabla T$.
Therefore, the Nernst signal $\vartheta_{yx}$,
defined as the {\em transverse} voltage $E_y$ generated by a thermal gradient $\nabla_x T$,
reads
\begin{equation}
%\vartheta_{xy}=-\frac{\sigma_{yy}\alpha_{xy}-\sigma_{xy}\alpha_{yy}}{\sigma_{xx}\sigma_{yy}-\sigma_{xy}\sigma_{yx}},~
\vartheta_{yx}=-\frac{\sigma_{xx}\alpha_{yx}-\sigma_{yx}\alpha_{xx}}{\sigma_{xx}\sigma_{yy}-\sigma_{xy}\sigma_{yx}}
\label{nernstsignal}
\end{equation}
and $\vartheta_{xy}$ is obtained from $x\leftrightarrow y$.
For a magnetic field $\vec B = B\hat z$ in $z$ direction,
the Nernst { \it coefficient \/} is usually defined as $\nu_{yx}=\vartheta_{yx}/B$,
which tends to become field-independent at small $B$.
We employ a sign convention such that the vortex Nernst coefficient is always positive
(formally $\nu_{xy}=-\vartheta_{xy}/B$, Ref.~\onlinecite{signnote}).
In general, the Nernst coefficient can be negative or positive, for example if it is caused by the
flow of charged quasiparticles.

The quasiparticle Nernst effect in metals in often small, which can be rationalized by
the so-called Sondheimer cancellation:\cite{sondheimer1948} For a single parabolic band,
the two contributions to $\vartheta$ \eqref{nernstsignal} exactly balance each other.
Generally, in any realistic system, such a cancellation will be incomplete. In systems
with a small (effective) Fermi energy the resulting Nernst signal can be
large.\cite{nernst_fl} This consideration is likely of relevance for states with
translational symmetry breaking where band backfolding produces small Fermi pockets,
although Eq.~\eqref{nernstsignal} shows that, in the presence of multiple quasiparticle
bands, the Nernst signal is not simply a superposition of the signals from the individual
bands (even if this applies to the elements of the $\hat\sigma$ and $\hat\alpha$
tensors). Together, this makes clear that magnitude (and sign) of the quasiparticle
Nernst signal in general depend on various microscopic details.

In systems with broken tetragonal symmetry, it is worth discussing the anisotropy of the
Nernst signal, i.e., the difference between $\vartheta_{yx}$ and $\vartheta_{xy}$,
Eq.~\eqref{nernstsignal}. Assuming Onsager reciprocity, the Hall conductivities obey
$\sigma_{xy} = -\sigma_{yx}$ independent of the crystal symmetry. Such a relation does
not hold for $\alpha_{xy,yx}$ in general, however, in the low-$T$ limit the Mott relation
can be derived from Boltzmann theory,
\begin{equation}
\alpha_{ij}=-\frac{\pi^2}{3} \frac{k_B^2T}{e} \frac{\partial \sigma_{ij}}{\partial \mu} |_{E_F} \ ,
\label{mottrelation}
\end{equation}
implying that $\alpha_{xy}
= -\alpha_{yx}$. Then, anisotropies in the Nernst signal can only arise from diagonal
conductivity anisotropies, $\sigma_{xx} - \sigma_{yy} \neq 0$ and $\alpha_{xx} -
\alpha_{yy} \neq 0$.

Strictly, Onsager reciprocity can only be proven if time reversal symmetry is not
spontaneously broken by the system. It has recently been proposed that a particular type
of time-reversal symmetry breaking is at play in underdoped cuprates, leading to
off-diagonal conductivity anisotropies.\cite{varma10} Here, we shall restrict our
attention to stripe (and related) states. Although magnetically ordered stripe states do
break time reversal as well, such states do not lead to off-diagonal anisotropies within
the Boltzmann framework described below in Sec.~\ref{sec:transp}, i.e., the above
statements based on Onsager reciprocity do apply.

%%%%%%%%%%%%%%%%%%%%%%%%%%%%%%%%%%%%%%%%%%%%%%%%%%%%%%%%%%%%%%%%%%%%%%%%%%%%%%%

\section{Quasiparticle model and Boltzmann transport}
\label{sec:model}

To calculate the normal-state quasiparticle Nernst effect,
we consider electrons moving on a square lattice of unit lattice constant,
with the two-dimensional dispersion given by
\begin{eqnarray}
\varepsilon_{  k} = &-& 2 t_1 (\cos k_x +\cos k_y ) - 4 t_2 \cos k_x \cos k_y \nonumber\\
                &-&2 t_3 (\cos 2 k_x +\cos 2 k_y) \ .
\label{dispersion}
\end{eqnarray}
The hopping parameters are chosen as $t_1=0.38$\,eV, $t_2=-0.32 t_1$
and $t_3=-0.5 t_2$,\cite{norman94,Andersen1995} unless otherwise noted.
The two-dimensional electron density is $n=1-x$ per unit cell.

\subsection{Mean-field theory for ordered states}

As is standard and is described in much detail in Refs.~\onlinecite{millis07,linmillis,hackl09b,hackl10},
we shall employ simple (non-self-consistent) mean-field approximations to capture the
symmetry-broken states.

Electron-nematic order is captured by different hopping energies along the $x$ and
$y$ axes. Hence, $t_{1x,y} = (1\pm\e/2) t_1$ and $t_{3x,y} = (1\pm\e/2) t_3$.

For stripes, the quasiparticles with the
dispersion~\eqref{dispersion} are subject to a periodic modulation in the site chemical
potential or bond kinetic energy.
The relevant modulations are described by
\begin{equation}
\hat{V}_1= \sum_{{\vec k},\sigma} \bigl( V_c({\vec k}) c_{{\vec k}+{\vec Q}_c\sigma}^\dagger c_{{\vec k}\sigma}^\pdag
+h.c. \bigr)
\label{vcdw}
\end{equation}
for a charge density wave and
\begin{equation}
\hat{V}_2= \sum_{\bk,\sigma} \sigma \bigl( V_s(\bk) c_{\bk+{\vec Q}_s \sigma}^\dagger c_{{\vec k}\sigma}^\pdag
+h.c. \bigr)
\label{vsdw}
\end{equation}
for a collinear spin density wave with polarization in $z$ direction.
In general, both $V_c({\vec k})$ and $V_s({\vec k})$ are complex, with
the phase corresponding to a sliding degree of freedom.
A site-centered CDW has modulated on-site chemical potentials,
and we choose a real $V_c({\vec k}) \equiv -V_c$.
A bond-centered CDW with on-site modulations is characterized by $V_c( \bk) \equiv -V_c
e^{-i Q_c/2}$; for modulations in the kinetic energy\cite{vs99,vojta2008} with primarily $d$-wave form factor
we have $V_c( \bk) = -\delta t (\cos(k_x+\frac{Q_c}{2})-\cos k_y ) e^{-iQ_c/2}$;
in both cases ${\vec Q}_c = (Q_c,0)$.
A site-centered SDW has again a real $V_s({\vec k}) \equiv V_s$,
whereas a bond-centered SDW is captured by $V_s(\bk) \equiv -V_s
(1 + e^{-iQ_c/2})/(2 \cos(Q_c/4))$ where ${\vec Q}_s = (\pi\pm Q_c/2,\pi)$.
The complex phases of the CDW and SDW mean fields
have been chosen such that the resulting order parameters $\phi_c$ and $\phi_s^2$
are in-phase. Moreover, with positive $V_c$ (site-centered) and positive $\delta t$
(bond-centered) the resulting modulations are such that the electron density is
small where the magnitude of the magnetic moment is small (i.e. near the anti-phase
domain walls).

Various forms of $V_c({\vec k})$ and $V_s({\vec k})$ were used in
Ref.~\onlinecite{hackl10}; below we will show results for site-centered SDW modulations.
Note that, on symmetry grounds, a mean-field Hamiltonian with a collinear SDW
modulation only will induce a (parasitic) CDW with ${\vec Q}_c=2{\vec Q}_s$.

\subsection{Semiclassical transport}
\label{sec:transp}

In the following, we shall calculate the low-temperature DC transport properties using a
Boltzmann equation. We further assume, as is appropriate for low temperatures, that the
relaxation is mainly due to randomly distributed impurities with a low density, leading
to a constant relaxation time $\tau_{\vec k}\equiv \tau_0$. To lowest order in the
applied fields, the conductivities required to calculate the Nernst signal are given by
\begin{eqnarray}
\alpha_{xx} &=&  \frac{2e}{T} \sum_{{\vec k}, n} \frac{\partial f_{\vec k}^0}{\partial \varepsilon_n({\vec k})}
\varepsilon_n({\vec k})  \tau_0 (v_{\vec k}^x)^2    \nonumber\\
\alpha_{xy} &=&  \frac{2e^2B}{T\hbar c} \sum_{{\vec k}, n} \frac{\partial f_{\vec k}^0}{\partial \varepsilon_n({\vec k})}
\varepsilon_n({\vec k}) \tau_0^2 v_{\vec k}^x
\biggl[ v_{\vec k}^y \frac{\partial v_{\vec k}^y}{\partial k_x} -  v_{\vec k}^x \frac{\partial v_{\vec k}^y}{\partial k_y}\biggr]\nonumber\\
\sigma_{xx} &=&  -2e^2 \sum_{{\vec k}, n} \frac{\partial f_{\vec k}^0}{\partial \varepsilon_n({\vec k})}
\tau_0 (v_{\vec k}^x)^2  \nonumber\\
\sigma_{xy} &=& -2 \frac{e^3 B}{\hbar c} \sum_{{\vec k}, n} \frac{\partial f_{\vec k}^0}{\partial \varepsilon_n({\vec k})}
\tau_0^2 v_{\vec k}^x  \biggl[ v_{\vec k}^y \frac{\partial v_{\vec k}^y}{\partial k_x} -  v_{\vec k}^x \frac{\partial v_{\vec k}^y}{\partial
k_y}\biggr],
\label{boltzmann}
\end{eqnarray}
where $\sum_n$ runs over the quasiparticle bands of the system in the presence of the
symmetry-breaking order, and $\varepsilon(\vec k)$, $v_{\vec k}$ are the corresponding
quasiparticle energies and velocities. We note that, in general, the current operators
have inter-band contributions. However, those do not show up in the semiclassical
Boltzmann equation, as the spectral functions appearing in the more general quantum
Boltzmann equation are replaced by $\delta$ functions at the quasiparticle energies in
the semiclassical limit. This approximation is justified for DC transport, provided that
coherent quasiparticles exist. As a result, Eqs.~\eqref{boltzmann} are diagonal in the
band index. A detailed derivation of Eqs.~\eqref{boltzmann} can be found, e.g., in
Ref.~\onlinecite{hackl10}.

Within the Boltzmann framework and in the low-temperature limit, the Nernst coefficient
$\nu$, is proportional to both the temperature $T$ and the relaxation time $\tau_0$.

%%%%%%%%%%%%%%%%%%%%%%%%%%%%%%%%%%%%%%%%%%%%%%%%%%%%%%%%%%%%%%%%%%%%%%%%%%%%%%%

\section{Quasiparticle Nernst effect: Results}
\label{sec:results}

In the remainder of the paper, we discuss the effect of the Fermi surface reconstruction
due to symmetry breaking on the normal-state Nernst effect in cuprates. Our focus will be
on the underdoped regime. Here, both nematic and stripe order have found experimental
support. While stripes have been established to exist in cuprates of the 214
family,\cite{stripe_rev2} the situation in YBCO is more involved: Both neutron
scattering\cite{hinkov08} and transport\cite{taill10b} experiments have been interpreted
in terms of electron-nematic order, setting in somewhere below the pseudogap line
$T^\ast$, with static stripe order being absent in this regime. On the other hand,
theoretical descriptions of Hall and quantum oscillation data using SDW mean-field
pictures have been found to match experiments relatively
well,\cite{millis07,linmillis,harrison09,millis10} suggesting that SDW (or stripe) order is at least
present in large magnetic fields of 15\,T and above.
Such field-induced order has been detected in neutron scattering in both 214 and YBCO
cuprates.\cite{stripe_rev2,chang09,haug_mf}
Theoretically, the main driving mechanism is anticipated to be the competition between
superconductivity and density-wave order.\cite{demler1}

\subsection{Stripes and Lifshitz transitions}
\label{sec:p12}

As pointed out above, the Fermi-surface reconstruction into pockets and open orbits will
crucially influence the Nernst signal. A particularly interesting situation was
advertised in Ref.~\onlinecite{millis10} for YBCO: it was argued that the electron
pockets near $(\pi,0)$, which have been made responsible for both the negative Hall
effect\cite{LeBoeuf07} and the dominant quantum oscillations with 530\,T oscillation
frequency,\cite{doiron07,sebastian08,aud09} disappear with decreasing doping, as a result
of a strong stripe modulation in combination with a large spatial modulation period. The
Lifshitz transition from a state with $(\pi,0)$ pockets to a state without $(\pi,0)$
pockets was proposed to be the origin of the divergence of the cyclotron mass upon
lowering doping, as is observed in quantum oscillation measurements.\cite{sebastian10b}

\begin{figure*}
\includegraphics[width=15cm]{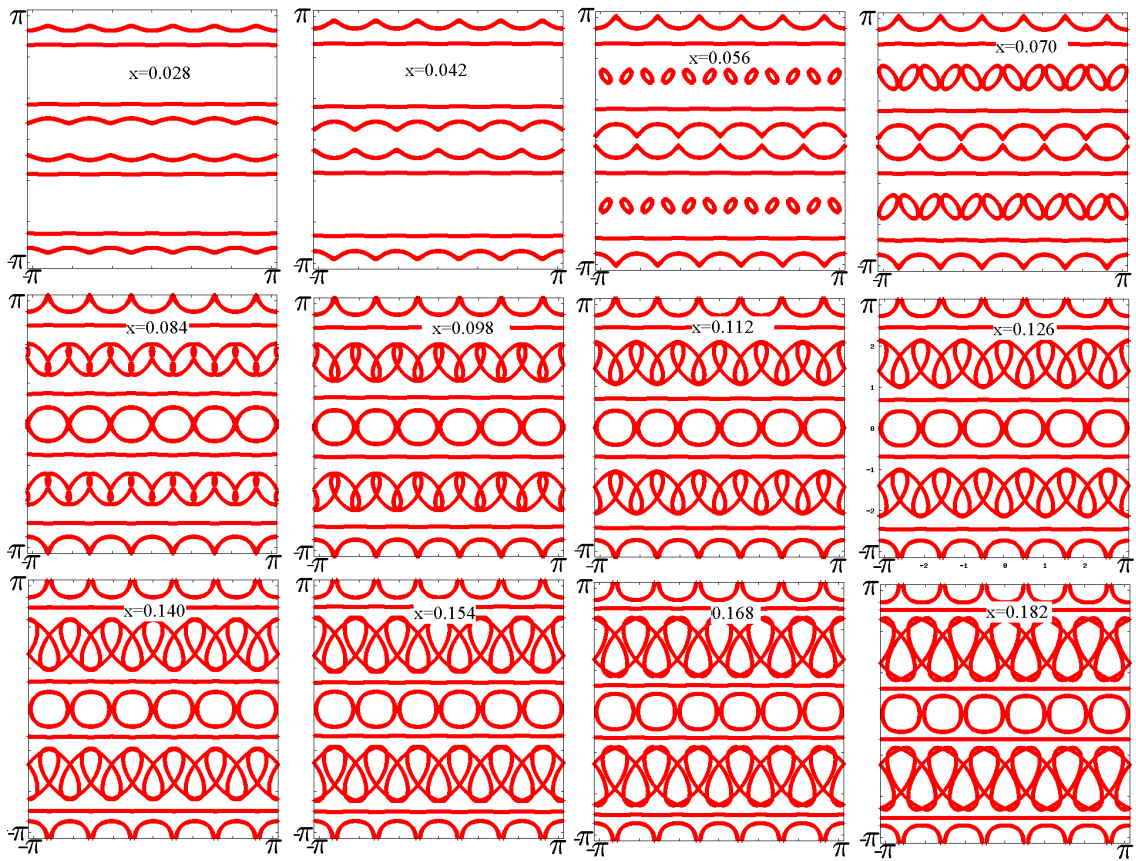}\\
\caption{
Fermi surfaces for vertical stripes with real-space period $M=12$ for different doping
values $x$. The bare quasiparticles, Eq.~\eqref{dispersion}, are subject to a
site-centered modulation in the spin sector, Eq.~\eqref{vsdw}, with a doping-dependent
strength $V_s(x) = (0.328 - x)$\,eV (Ref.~\onlinecite{millis10}). Various Lifshitz
transitions are visible; in particular, the electron pockets near $(\pi,0)$ merge into
open orbits for $x\lesssim 8.5\%$.
}
\label{fig:fs1}
\end{figure*}

Fig.~\ref{fig:fs1} shows a sequence of Fermi surface plots for a (vertically striped) SDW
state with real-space period $M=12$ (implying a CDW period of $N=6$) and varying doping.
Thereby, it is assumed that the modulation strength follows $V_s(x) = (0.328 - x)$\,eV.
Note that, from experiments on 214 compounds,\cite{jt95,yamada98,stripe_rev1,antonio_rev,stripe_rev2}
$M=12$ is known to be appropriate for doping $x\approx 1/12$; in
Ref.~\onlinecite{millis10} $M=12$ was in fact employed, together with the above parametrization $V_s(x)$,
to match quantum oscillation data over the entire doping range between 9\% and 15\%.
(Clearly, $M=12$ and $V_s(x)$ cease to be relevant to cuprates outside this doping
window.)
A more detailed modeling requires to take into account the doping dependence of $M$
(which moreover may depend on the applied field $B$ as well). Unfortunately, $M(x,B)$ is
not well known for YBCO, therefore we restrict our attention to $M=12$ which we believe
to reproduce important qualitative aspects.

Technically, $M=12$ implies the diagonalization of a $12 \times 12$ Hamiltonian matrix,
in analogy to Refs.~\onlinecite{linmillis,hackl10}. We consider a site-centered spin-only
modulation with $V_s({\vec k}) = V_s$, unless otherwise noted.

The Fermi surfaces in Fig.~\ref{fig:fs1} make clear that a series of Lifshitz transitions
can be expected as doping is varied. At dopings $x\geq 8.5\%$, various types of Fermi
pockets are present, in particular there are prominent electron pockets near $(0,\pi)$
and equivalent wavevectors. Those electron pockets merge and disappear for $x\lesssim
8.5\%$, leaving only small electron pockets and open orbits. At around $7.5\%$, these small
hole pockets transform into hole pockets, until, finally, for $x\lesssim5\%$ all pockets
merge, and the resulting state displays exclusively open orbits. The presence of $(\pi,0)$
electron pockets at dopings of 10\% and higher is a robust feature of spin stripe order,
see e.g. Fig.~2 of Ref.~\onlinecite{millis10} and Figs.~2,3 of Ref.~\onlinecite{hackl10} for
Fermi-surface results for $M=8$ stripes.\cite{pocket_foot} In contrast, additional strong
charge modulation may eliminate the electron pockets, Fig.~4b of Ref.~\onlinecite{hackl10}.

\subsection{Nernst effect from open orbits}
\label{sec:open}

Before discussing the Nernst effect over the entire doping range, we find it appropriate
to separately analyze the Nernst signal arising from open Fermi orbits. Clearly, those
will dominate in stripe-ordered cuprates at small doping, but the information here is
also interesting on general grounds.

\begin{figure}
\includegraphics[width=7.5cm]{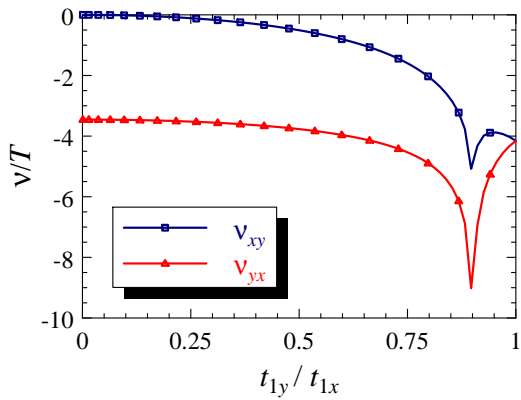}
\caption{
Nernst effect for a single band on a square lattice with anisotropic hopping.
For $t_{1y}/t_{1x} < 0.9$, the Fermi surface consists of open orbits running along the
$\hat y$ direction. For small $t_{1y}$, the Nernst signal develops a huge anisotropy.
The units of $\nu/T$ are
$10^{5} {\rm V} / ({\rm K}^2 {\rm T}) \times \tau_0/{\rm s}$;
here $t_{1x} = 1$\,eV.
}
\label{fig:open}
\end{figure}

To illustrate the physics, we consider a simple tight-binding model on a square lattice,
with nearest-neighbor hopping in $\hat{x}$ direction, $t_{1x}$ normalized to unity. We fix the
hole doping at $x=0.1$ and vary the nearest-neighbor hopping in $\hat{y}$ direction, $t_{1y}$.
Both Nernst signals, $\nu_{yx}/T$ and $\nu_{xy}/T$, are displayed in Fig.~\ref{fig:open} as
function of $t_{1y}/t_{1x}$. A van-Hove singularity occurs at $t_{1y}/t_{1x} \approx 0.9$
where the electron-like Fermi surface of the isotropic limit changes into open orbits. At
the van-Hove singularity both Nernst signals are enhanced. Interestingly, in the
open-orbit regime, $\nu_{yx}/T$ remains finite, while $\nu_{xy}/T$ tends to zero as
$t_{1y}\to 0$. This implies a huge anisotropy of the Nernst signal for open
orbits.\cite{open_foot} Recall that $\nu_{yx}$ is Nernst signal for $\vec{\nabla} T
\parallel \hat x$, i.e., a strong Nernst signal occurs for $\vec{\nabla} T$ ($\vec E$)
parallel (perpendicular) to the strong hopping direction.
We note that the same information was already present in Fig.~3b of
Ref.~\onlinecite{hackl09b}, where a large hopping anisotropy $\epsilon$ in an
electron-nematic state led to a Lifshitz transition to a Fermi surface with open orbits,
resulting in $|\nu_{xy}| \ll |\nu_{yx}|$ as in Fig.~\ref{fig:open}.

The limit of small $t_{1y}$ can be understood analytically. Power counting in the
Boltzmann expressions for the conductivities shows that $\sigma_{xx},\alpha_{xx} \propto
t_{1y}^0$, while the other quantities $\sigma_{yy}$, $\sigma_{xy}$, $\alpha_{yy}$,
$\alpha_{xy}$ scale as $t_{1y}^2$. From Eq.~\eqref{nernstsignal} it follows that
$\nu_{yx} \propto t_{1y}^0$ while $\nu_{xy} \propto t_{1y}^2$, consistent with
Fig.~\ref{fig:open}. Note that, of course, there is no transverse response for $t_{1y} =
0$, i.e., there is no meaningful definition of the Nernst signal in the strict
one-dimensional (1d) limit.
Finally, the sign of $\nu_{yx}$ for small $t_{1y}$ is not robust, but depends on details
of the quasi-1d band structure.

\subsection{Nernst effect from stripes in underdoped cuprates}

\begin{figure}
\includegraphics[width=8cm]{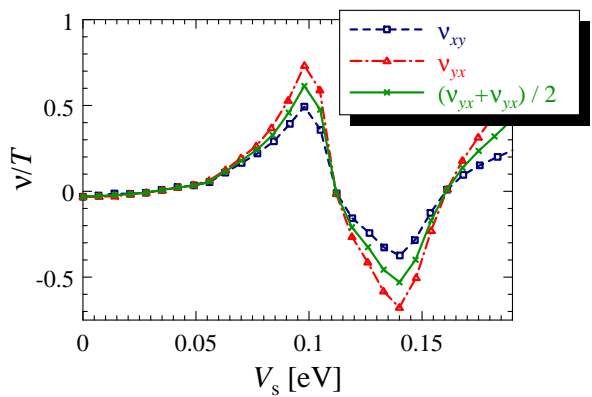}
\caption{\label{fig:p8}
Nernst effect for period-8 antiferromagnetic stripes at doping $x=1/8$ as function of
the spin modulation. The bare quasiparticle dispersion is in Eq.~\eqref{dispersion}.
The Nernst coefficient becomes negative at $V_s \simeq 0.1$\,eV, corresponding to maximal
local moments of $2\mu_B \langle S_z \rangle \simeq 0.3 \mu_B$.
Here, $\nu_{yx}$ is Nernst signal for $\vec{\nabla} T \parallel \hat x$.
The stripes have a modulation wavevector $\parallel \hat x$, i.e., run along $\hat y$,
such that $\nu_{xy}$ ($\nu_{yx}$) is defined with $\vec{\nabla} T$ parallel
(perpendicular) to the stripes.
(Figure adapted from Ref.~\onlinecite{hackl10})
}
\end{figure}

Let us start by summarizing the main results obtained previously in
Ref.~\onlinecite{hackl10}. There, mainly stripes with period $M=8$ and doping levels of
1/8 and higher have been considered. For spin stripes with realistic amplitudes of the
ordered magnetic moments, $2\mu_B \langle S_z \rangle < 0.3 \mu_B$, the Nernst signal was
found to be strongly enhanced, with its sign being positive. The result is reproduced in
Fig.~\ref{fig:p8}. Depending on the admixture of charge modulations, sign changes in the
Nernst signal occurred as function of modulation strength or doping. Some calculations
were also performed for modulation periods $M=10$ and 16 appropriate for more underdoped
cuprates, with qualitatively similar results, but a robust positive Nernst signal
remained restricted to period-8 spin stripes.

For most stripe states investigated in Ref.~\onlinecite{hackl10}, the anisotropy of the
Nernst signal was found to be weak. This can be interpreted as arising from the averaging
over various anisotropic orbits of the backfolded band structure. This weak anisotropy is
in contrast to the huge anisotropy which results from the orthorhombic distortion of a
single large Fermi surface sheet in an electron-nematic state.\cite{hackl09b} There, a
subtle interference of the diagonal anisotropies in the transport tensors $\hat\sigma$
and $\hat\alpha$ can cause a large Nernst anisotropy even if the anisotropies in $\sigma$
and $\alpha$ are moderate.

\begin{figure}
\includegraphics[width=8cm]{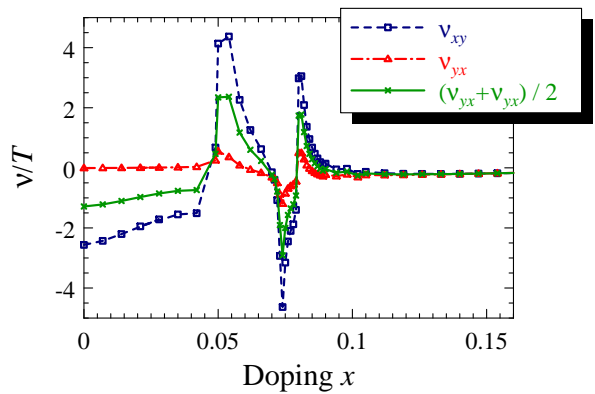}
\caption{
Nernst signal as in Fig.~\ref{fig:p8}, but now for period-12 antiferromagnetic stripes as
function of doping $x$. The modulation strength is chosen as $V_s(x) = (0.328 - x)$\,eV.
The regime of open orbits, $x<5\%$, is characterized by a strong Nernst anisotropy.
}
\label{fig:p12}
\end{figure}

We are now in the position to discuss the Nernst signal from spin stripes for strongly
underdoped cuprates as introduced in Sec.~\ref{sec:p12}. We again fix $M=12$, use
vertical stripes, and employ $V_s(x) = (0.328 - x)$\,eV as proposed in
Ref.~\onlinecite{millis10}. The resulting Nernst coefficients $\nu/T$ for both directions
are displayed in Fig.~\ref{fig:p12}.
Most strikingly, the signal is strongly enhanced at the various Lifshitz transitions at
$x\approx 5\%$, 7.5\%, and 8.5\%, see Sec.~\ref{sec:p12}. (Note that jumps in $\nu/T$ occur at
Lifshitz transitions where Fermi pockets open or close; this has been analyzed in some
detail in Ref.~\onlinecite{hackl09a}.)

At dopings $x>10\%$, the extended pockets lead to a small Nernst signal with little anisotropy. In the opposite
limit of small dopings $x<5\%$, only $\nu_{xy}$ is sizeable while $\nu_{yx}$ is tiny --
this is consistent with the discussion in Sec.~\ref{sec:open}, with the difference that
here the open Fermi orbits run in $\hat x$ direction, while they run along $\hat y$
for small $t_{1y}$ in Sec.~\ref{sec:open}.
The emergence of hole pockets at $x\approx 5\%$ then leads to a large positive Nernst
signal, which changes sign multiple times upon further increasing $x$ where electron
pockets emerge, Fig.~\ref{fig:fs1}. All features are much more pronounced in  $\nu_{xy}$
as compared to $\nu_{yx}$, reflecting general the quasi-1d character of the Fermi
surfaces.

%%%%%%%%%%%%%%%%%%%%%%%%%%%%%%%%%%%%%%%%%%%%%%%%%%%%%%%%%%%%%%%%%%%%%%%%%%%%%%%

\section{Discussion}

In this paper, we have summarized\cite{hackl09b,hackl10} and extended the analysis of the
Nernst signal from quasiparticles in cuprates with broken lattice symmetries. For stripe
states, we find the quasiparticle Nernst signal $\nu/T$ to be significantly enhanced
(compared to the unmodulated state), which can be traced back to the presence of small
Fermi pockets. The sign of $\nu/T$ depends on the period and strength of the modulation
and other microscopic details, with a positive $\nu/T$ being found for period-8 spin
stripes. In the small-doping regime, where the Fermi surface of the stripe state consists
of open orbits only, the Nernst signal in one direction is tiny, while in the other it is
strongly negative. The latter is in fact the only case where a robust and large
anisotropy in the stripe-induced Nernst signal was found. Note that for a nematic
distortion of a single Fermi surface sheet, the Nernst anisotropy is robustly large,
in particular near a van-Hove singularity.

\subsection{Robustness}

As our calculations involved a series of approximations, one may ask which features can
be expected to be robust. The two most drastic approximations are probably (i) the
mean-field approximation for the ordered states and (ii) the relaxation-time
approximation with constant $\tau_0$.

At present, it is difficult to estimate the effects of order-parameter fluctuations
beyond mean field (i). However, a better understanding of those might be needed in
particular in the context of fluctuating stripes, which have been suggested to occur at
elevated temperatures in various cuprates. In contrast, deep in the ordered state, we can
assume the mean-field approximation to be qualitatively justified.

Approximation (ii) neglects both the temperature and energy dependence and the anisotropy
of the relaxation rate. The angular dependence of $\tau_0$ may well be important
for the Nernst response of stripe states, because all conductivities receive
contributions from the various Fermi pockets, which will in general have different
relaxation rates. Hence, the enhancement of the Nernst signal near van-Hove singularities
and Lifshitz transitions can be expected to be robust, but the sign of the Nernst signal away
from these singularities may change upon including anisotropic scattering.
Importantly, our qualitative conclusions for Nernst {\em anisotropies} are rather robust: For
instance, for the electron-nematic order discussed in Ref.~\onlinecite{hackl09b}, only a
strong difference in scattering rates at $(\pi,0)$ and $(0,\pi)$ could spoil the result
which is unlikely.

\subsection{Experiments}

Nernst effect investigations in 214 cuprates have an extended history, and the enhanced positive
Nernst signal at intermediate temperatures above $\Tc$ has been commonly interpreted as
signature of vortex physics, i.e., preformed Cooper pairs.\cite{xu00,wang06} The recent
identification\cite{taill09} of an additional positive contribution, whose temperature
dependence appears to track some characteristic pseudogap (or charge order) scale, has
been qualitatively verified by others,\cite{hess10} but its interpretation\cite{taill09}
in terms of a Fermi-surface reconstruction due to charge order has been
questioned.\cite{li10,hess10}

Our results, showing a positive Nernst signal for period-8 stripes,\cite{hackl10} are in principle
consistent with this interpretation. However, the fact that only minor differences in the
Nernst signal exist between stripe-ordered \eulsco\ and \ndlsco\ on the one hand and
non-stripe-ordered \lsco\ on the other hand\cite{taill09,hess10} complicates matters.
As \lsco\ is expected to have strong stripe {\em fluctuations}, a scenario of fluctuating
stripes\cite{stripe_rev1,vvk} being responsible for the Nernst enhancement is viable, but a concrete
theoretical analysis (which also has to account for pinning of stripes by defects)
is lacking.

For underdoped YBCO, the Nernst signal is negative and strongly anisotropic in a
doping-dependent window of intermediate temperatures.\cite{taill10b} This appears to be
consistent with the results of Ref.~\onlinecite{hackl09b}, supporting the interpretation
that electron-nematic order in at play in YBCO (with the built-in orthorhombicity acting
as a field which smears the nematic phase transition and aligns the domains).

Nernst measurements on more underdoped YBCO should be extremely interesting, in
particular in a regime where field-induced incommensurate magnetism has been
detected\cite{haug_mf} and where the superconducting $\Tc$ is small. Then, one can expect
that our results for long-period spin stripes become relevant. Based on
Fig.~\ref{fig:p12}, we predict that at low doping also the low-temperature Nernst signal
should be strongly anisotropic, with $\nu_{yx}$ ($\nu_a$ in the terminology of
Ref.~\onlinecite{taill10b}) being tiny (assuming that the stripes run along the
orthorhombic b axis\cite{stripe_rev2,hinkov08}).

%%%%%%%%%%%%%%%%%%%%%%%%%%%%%%%%%%%%%%%%%%%%%%%%%%%%%%%%%%%%%%%%%%%%%%%%%%%%%%%

\begin{acknowledgments}
We acknowledge useful discussions as well collaborations on related subjects
with L.~Fritz, A. Rosch, S. Sachdev, L.~Taillefer, and A.~Wollny.
The research has been supported by the DFG through the SFB 608 (K\"oln)
and the Research Units FOR 538 and FOR 960.
\end{acknowledgments}

%%%%%%%%%%%%%%%%%%%%%%%%%%%%%%%%%%%%%%%%%%%%%%%%%%%%%%%%%%%%%%%%%%%%%%%%%%%%%%%


\begin{thebibliography}{99}
\bibitem{signnote}
For fixed magnetic field $\vec B || \hat z$,
a $C_4$-symmetric situation yields $\vartheta_{xy} = -\vartheta_{yx}$ from
Eqs.~(\ref{thermoelectrics},\ref{nernstsignal}).
Here we use the convention
that the three vectors $\vec E$, $\vec\nabla T$, and $\vec B$ form a right-handed
system for the measurements of both $\nu_{xy}$ and $\nu_{yx}$.
Then, $\nu_{xy} = \nu_{yx}$ for $C_4$ symmetry.
\bibitem{pocket_foot}
In Ref.~\onlinecite{hackl10}, the identification of pockets as electron-like or hole-like
was partially incorrect. This concerns the captions of Figs.~2b, 3, and 14. All numerical
remain are unaffected.
\bibitem{open_foot}
Ref.~\onlinecite{hackl10} stated that open orbits do not appreciably contribute to the Nernst
signal. While true for some of the stripe states investigated in Ref.~\onlinecite{hackl10}, this is not
generally valid, see Sec.~\ref{sec:open} of the present paper.
\end{thebibliography}
\end{document}